\documentclass[%
 aip,
 jmp,%
 amsmath,amssymb,
reprint,%
]{revtex4-1}

\usepackage{graphicx}
\usepackage{dcolumn}
\usepackage{bm}
\usepackage{color}
\begin{document}

\preprint{}

\title{Transmission of spin waves in ordered FeRh epitaxial thin films}



\author{Takamasa Usami}
\author{Ippei Suzuki }
\author{Mitsuru Itoh}
\author{Tomoyasu Taniyama}
\email{taniyama.t.aa@m.titech.ac.jp}
\affiliation{Laboratory for Materials and Structures, Tokyo Institute of Technology, 4259 Nagatsuta, Yokohama 226-8503, Japan}


\date{\today}

\begin{abstract}
We report on B2-ordering dependence of magnetostatic surface spin waves in ferromagnetic FeRh at room temperature. Spin waves transmit over a distance longer than 21 $\mu$m in highly ordered FeRh alloys even with relatively large spin-orbit interaction. The long-range transmission likely arises from the induced Rh moments of the ordered FeRh due to ferromagnetic exchange interaction between Fe and Rh. The results indicate a potential of using FeRh in spintronic and magnonic applications by integrating with other fascinating magnetic characteristics of FeRh such as electric field induced magnetic phase transition.
\end{abstract}


\maketitle 

Spin waves or magnons, that is, well-defined collective propagation modes of the precession of magnetic moments in a ferromagnet, have attracted great attention due to their potential for use in information transmission media\cite{Schneider,Klingler,Bailleul,Kiseki,Yamanoi}. The less energy-dissipative propagation of spin waves in ferromagnets provides the breakthrough that opens up a new pathway for exploiting low-power electronic technologies. In this perspective, spin wave propagation in ferrimagnetic insulator yttrium iron garnet\cite{Schneider,Klingler} and metallic NiFe\cite{Bailleul,Kiseki,Yamanoi} that show very low damping of spin precession during the propagation has been investigated. Spin waves in other magnetic materials, however, have little been discussed although a number of magnetic materials have already been utilized for magnetic and spintronic applications so far.

Among a variety of magnetic materials, unique magnetic properties of B2-ordered FeRh alloys are currently being researched extensively, e.g., antiferromagnetic (AFM) - ferromagnetic (FM) phase transition\cite{Fallot,Maat,Vries,Baldasseroni,Bordel,Thiele,Gray,Kouvel,Lommel,Barua,Suzuki1,Suzuki2} accompanied by a large reduction in the resistivity\cite{Lommel}, an isotropic volume expansion of $\sim$1\% at the AFM-FM phase transition\cite{Kouvel}, giant magneto-resistance\cite{Suzuki2}, laser driven ultrafast switching of the magnetic phases\cite{Hohlfeld}, and exchange bias at AFM FeRh/ferromagnetic metal interface\cite{Suzuki4,Yamada2}. Recent work also has demonstrated electric field induced AFM-FM phase transition due to strain transfer effects\cite{Suzuki3,Cherifi,Phillips,Taniyama}, spin polarized current induced AFM-FM phase transition in FeRh\cite{Naito,Suzuki5}, and AFM memristers\cite{Marti,Moriyama}. Such exciting magnetic properties offer an opportunity to be integrated in a vastly expanded range of spintronic applications. Despite, spin waves in FeRh has never been explored experimentally; even excitation and detection of spin waves have not been reported.

In this paper, we report on excitation and detection of magnetostatic surface spin waves in ferromagnetic FeRh with different B2-ordering. B2-order parameter ($S$) dependence of the dispersion relationship of spin waves in FeRh epitaxial thin films is discussed. We find that the transmission length of the spin waves is relatively long over  21 $\mu$m in highly ordered FeRh with $S=0.75$. The long transmission length in the ordered FeRh is likely associated with the induced Rh moments by ferromagnetic exchange interaction between Fe and Rh in the ordered structure. From the fundamental physics points of view, the B2-ordering dependence of the spin wave characteristics could also provide a clue to the origin of the phase transition since the AFM-FM phase transition has close correlation with spin wave excitations as reported theoretically\cite{Ayuela,Gu,Sandratskii}.

\begin{figure}[t]
\includegraphics[width=7cm]{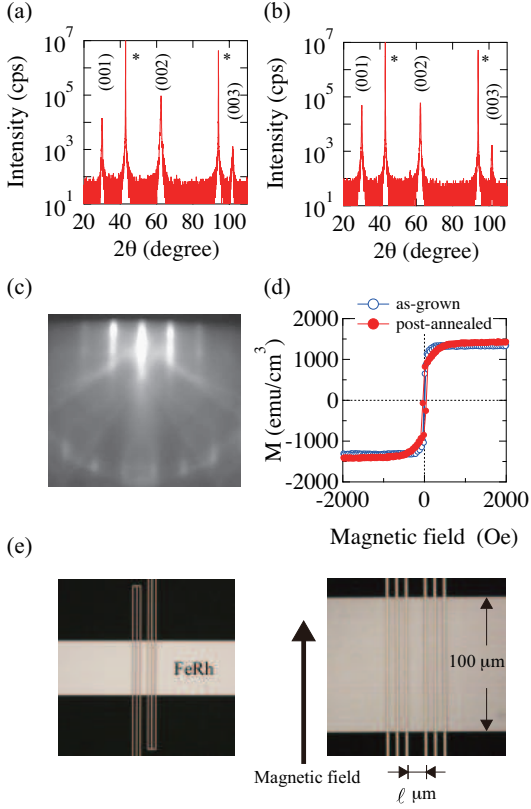}%
\caption{XRD patterns of (a) an as-grown FeRh thin film with $S=0.33$ and (b) the post-annealed FeRh thin film with $S=0.75$. (c) RHEED pattern of an FeRh thin film, (d) Magnetization curves of the two FeRh films in in-plane transverse magnetic field along MgO[100](001) direction, (e) Photograph of a device for magnetostatic surface spin wave measurement. The photograph on the right hand side is an expanded view near the co-planar wave guides.}
\label{Fig1}
\end{figure}

40-nm-think Fe$_{60}$Rh$_{40}$/MgO(001) epitaxial thin films were grown at 450$^{\circ}$C by co-evapo-ration of Fe and Rh in a molecular beam epitaxy (MBE) chamber with a base pressure under 10$^{-10}$ Torr. An as-grown film was cut into two pieces, and one of which was post-annealed at 620$^{\circ}$C in vacuum to obtain two FeRh thin films with different B2-order parameters. X-ray diffraction clearly shows the epitaxial growth (Figs. \ref{Fig1}(a) and (b)) and the relative integrated peak intensities of (001) and (002) allow to estimate the B2-order parameters $S$ of the films using Eq. (\ref{order_param})\cite{Vries}.
\begin{equation}
S=\sqrt{I_{001}^{\rm exp}/I_{002}^{\rm exp}}/\sqrt{I_{001}^{\rm cal}/I_{002}^{\rm cal}}\cong\sqrt{I_{001}^{\rm exp}/I_{002}^{\rm exp}}/1.07,
\label{order_param}
\end{equation}
where $I_{001}^{\rm exp}$, $I_{002}^{\rm exp}$, $I_{001}^{\rm cal}$, and $I_{002}^{\rm cal}$ are the experimentally obtained (001) and (002) diffraction intensities and the theoretical ones based on the B2 structure. 
The order parameters of the as-grown and post-annealed films are estimated to be $S=0.33$ and 0.75, respectively. Reflection high energy electron diffraction (RHEED) pattern also ensures the epitaxial growth and the flat surface as shown in Fig. \ref{Fig1}(c). 

Magnetization of the film was measured in in-plane magnetic fields along MgO[100](001) by using a vibrating sample magnetometer (VSM) at room temperature. Figure \ref{Fig1}(d) shows the magnetization curves of the as-grown and post-annealed films. The saturation magnetizations are 1310 emu/cm$^{3}$ and 1440 emu/cm$^{3}$, respectively. Note that the magnetic coercivity is larger for the post-annealed film than the as-grown film while the magnetization process of the post-annealed film exhibits a two-step magnetization process. The characteristic magnetization process for the post-annealed film is due to the crystalline magnetic anisotropy associated with the B2-ordering of Fe and Rh. 

FeRh epitaxial films were micro-patterned using electron beam lithography and Ar ion milling for spin wave excitation and detection measurements. Figure \ref{Fig1}(e) is a typical optical photograph of a spin wave device.  Au/Ti co-planar waveguides with ground-signal-ground (G-S-G) geometry are located on FeRh epitaxial films with an insulating SiO$_{x}$ interlayer. The distances $\ell$ between the two co-planar waveguides is designed to be 7 -- 21 $\mu$m as depicted in Fig. \ref{Fig1}(e). One of the co-planar waveguides is used to excite spin waves in the FeRh films and the reflection spectra $S_{11}$ and the transmission spectra $S_{21}$ are measured at room temperature using a vector network analyzer in in-plane magnetic fields transverse to the propagation direction of the spin waves up to 163 mT. The magnetic field geometry used allows to excite magnetostatic surface spin waves\cite{Kiseki,Yamanoi}.

\begin{figure}[t]
\includegraphics[width=7cm]{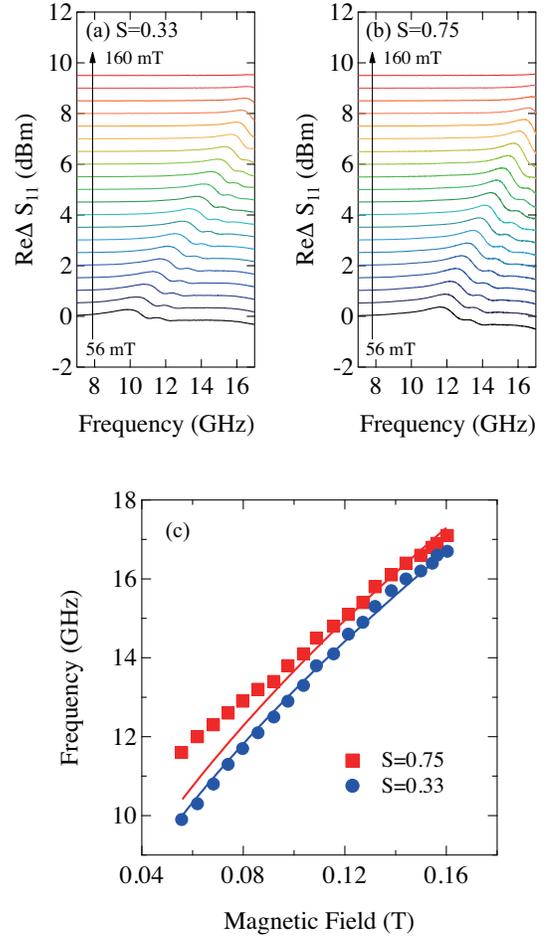}%
\caption{Reflection spectra $S_{11}$ measured at various magnetic fields for the FeRh thin films with (a) $S=0.33$ and (b) $S=0.75$. The distance between the co-planar waveguides is $\ell =14 \mu$m. (c) Dispersion relationships obtained from the $S_{11}$ spectra.}%
\label{Fig2}
\end{figure}

Figures \ref{Fig2} show the $S_{11}$ spectra of the as-grown and annealed FeRh thin films with different B2-order parameters $S=0.33$ and 0.75, respectively. 
Clear features associated with the magnetostatic surface spin waves are seen in the spectra. The spectra are composed of the fundamental and higher-order excitation features and the frequency at which the $S_{11}$ peaks shifts towards higher frequencies with increasing magnetic field. From Figs. \ref{Fig2}(a) and (b), the frequencies of the magnetostatic surface spin waves are plotted as a function of magnetic field in Fig. \ref{Fig2}(c). Eq. (\ref{dispersion_Eq}) that is applicable to magnetostatic surface spin waves can be fitted to the dispersion relationships\cite{Kiseki,Yamanoi}.
\begin{equation}
f=\frac{\gamma}{2\pi}\sqrt{B(B+M_{S})+\left(M_{S}/2\right)^2(1-e^{-2kd})},
\label{dispersion_Eq}
\end{equation}
where $\gamma$ is the gyromagnetic ratio, $B$ is the transverse in-plane magnetic field, $M_{S}$ is the saturation magnetization of the film, $k$ is the wave number, and $d$ is the film thickness. Since the wave number is defined by the distance $l=n\pi/k$ between the electrodes of the co-planar waveguides, we fix $k=n\pi /\ell$ (1/$\mu$m). It is clearly seen that the fitted curve well represents the $S_{11}$ of the as-grown film over the entire magnetic field region, providing $M_{S}=1.73$ T which is in agreement with the magnetization data in Fig. \ref{Fig1}(d). However, a fit deviates from the data points for the post-annealed film below 0.1 T while a fit above 0.1 T yields $M_{S}=1.86$ T, compatible with the magnetization data. The deviation is likely due to the additional magnetic crystalline anisotropy associated with the higher B2-ordering and the resultant tilting of the magnetization direction from the magnetic field direction. In fact, Fig. \ref{Fig1}(d) shows that the magnetization does not saturate until $\sim$0.1 T.

\begin{figure}[b]
\includegraphics[width=7cm]{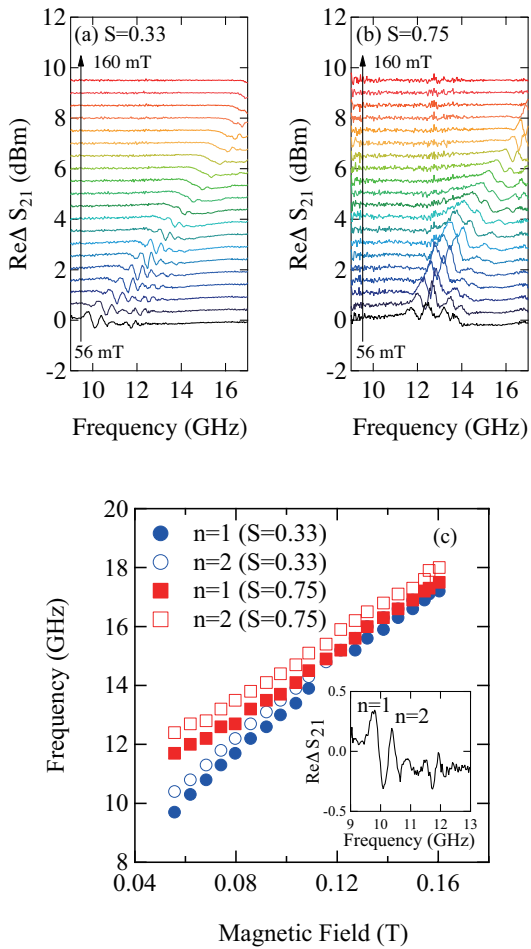}%
\caption{Transmission spectra $S_{21}$ measured at various magnetic fields for the FeRh thin films with (a) $S=0.33$ and (b) $S=0.75$. The distance between the co-planar waveguides is $\ell =14 \mu$m. (c) Dispersion relationships obtained from the fundamental ($n=1$) and second order ($n=2$) peaks in the $S_{21}$.}%
\label{Fig3}
\end{figure}

Shown in Fig. \ref{Fig3} is the transmission spectra $S_{21}$ measured at various in-plane magnetic fields for the two FeRh films. 
Similar to the reflection spectra $S_{11}$ in Fig. \ref{Fig2}, the $S_{21}$ spectra exhibit the oscillatory features which are more significant than that of $S_{11}$. The frequency at which the spectra peaks shifts toward a higher frequency regime with increasing magnetic field while the amplitude of the peaks is suppressed. Note that the amplitude of the peak features is more pronounced for the highly ordered film. Figure \ref{Fig3}(c) is the dispersion relationship of the transmission spectra $S_{21}$. The fundamental and the second order peaks which are assigned in the inset of Fig. \ref{Fig3} are plotted. Analogous to the dispersion relationship obtained from the $S_{11}$ spectra, the slope of the magnetic field dependence deviates from each other in the low field region. This is also likely due to the additional magnetocrystalline anisotropy in the highly B2-ordered film. The difference in the frequencies between $n=1$ and $n=2$ peaks does not change significantly over the entire field region. Since the difference in the frequency is related to the group velocity of the spin waves, i.e., $v_{g}=\Delta\omega/\Delta k=2\ell\Delta f$\cite{Yamanoi}, the group velocity in the FM FeRh can be estimated to be 11 $\mu$m/nsec and 17$\mu$m/nsec for the thin films with $S=0.33$ and 0.75, respectively. The group velocity of the magnetostatic surface spin waves $v_{g}=d\omega/dk$ is also calculated using Eq. (\ref{group_vel}), 
\begin{equation}
v_{g}=\frac{\gamma d\left(\displaystyle M_{s}/2\right)^{2}e^{-2kd}}{\sqrt{B(B+M_{s})+\left(\displaystyle M_{s}/2\right)^{2}(1-e^{-2kd})}}
\label{group_vel}
\end{equation}
providing $v_{g}=12 \mu$m/nsec and 14$\mu$m/nsec for the films, in good agreement with the values obtained from the dispersion relationships in Fig. \ref{Fig3}.

\begin{figure}[t]
\includegraphics[width=7cm]{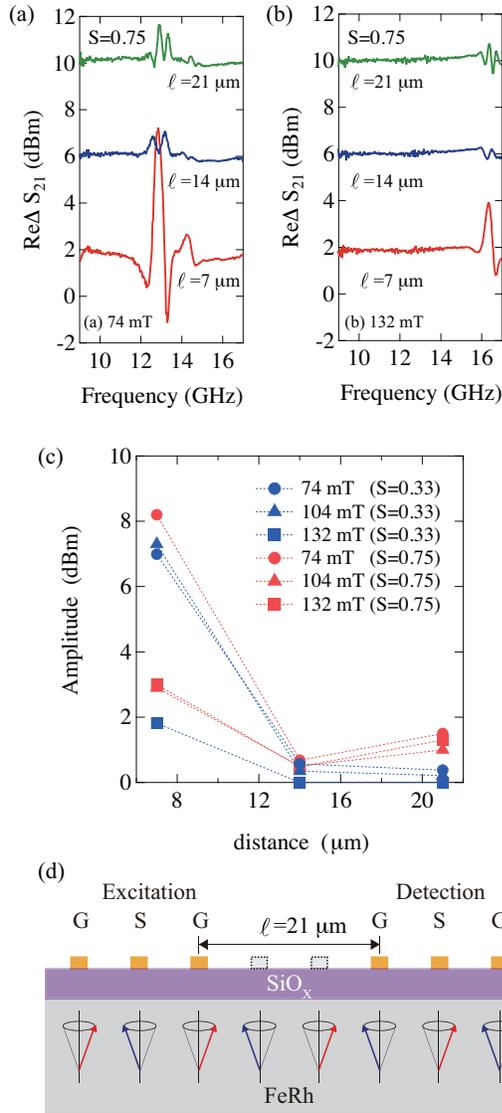}%
\caption{Transmission spectra measured at different electrode distance $\ell$ in magnetic fields of (a) 74 mT and (b) 132 mT. (c) The amplitude of the spin wave signals as a function of $\ell$. (d) Schematic illustration of spin waves excited in FeRh.}%
\label{Fig4}
\end{figure}

The electrode distance $\ell$ dependence of the transmission spectra $S_{21}$ also provide interesting aspects of the effect of the B2-ordering on the spin wave transmission. Figures \ref{Fig4}(a) and (b) show the $S_{21}$ spectra in magnetic fields of 74 and 132 mT for the annealed film recorded using the waveguides with $\ell$. 
Although the amplitude of the spin wave signal decreases with increasing $\ell$, the magnetostatic surface spin waves are still clearly observed even at $\ell$=21 $\mu$m. This is a striking result because 4$d$ Rh atoms with its large atomic number might causes a significant damping of the spin precession due to the large spin-orbit coupling. This interesting phenomena should be associated with possible Rh magnetic moments induced by Fe-Rh FM exchange coupling. As widely observed in FeRh experimentally as well as theoretically\cite{Lounis,Suzuki1}, Rh atoms possess magnetic moments in the FM state due to the exchange coupling, indicating that not only Fe moments but also Rh moments contribute to the transmission of spin waves in FeRh.

The transmission characteristics of spin waves for $S=0.75$ and $S=0.33$ also show clear B2-ordering dependence in Fig. \ref{Fig4}(c). The amplitude of spin wave signals in $S_{21}$ is more significant for the post-annealed film. We note that the spin wave signal for the higher ordered film ($S=0.75$) is seen even at $\ell =21 \mu$m while that in the film with $S=0.33$ almost disappears. This is also compatible with the description that the highly ordered FeRh possess larger Rh moments in the FM state due to the FM exchange interaction with Fe moments, leading to a longer spin wave transmission length. 

It should be noticed that the phase of the spin signals for $\ell =7$ and $21 \mu$m in Fig. \ref{Fig4}(a) differs from that for $\ell =14 \mu$m by 180$^{\circ}$, that is, the fundamental peak in the $S_{21}$ spectra for  $\ell =7$ and $21 \mu$m corresponds to the dip in the spectra for $\ell =14 \mu$m. The behavior is understood that the phase of magnetostatic spin waves for $\ell =7$ and $21 \mu$m is opposed to that for $\ell =14 \mu$m as depicted in Fig. \ref{Fig4}(d), further ensuring that magnetostatic surface spin waves are well defined by designing with the geometry of the co-planar waveguides used.

In summary, we have observed magnetostatic surface spin waves in FM FeRh thin films by using co-planar waveguides at room temperature. The spin wave signals show clear B2-ordering dependence, where higher ordering gives rise to a longer transmission length of spin waves. It is also found that the excited spin wave reaches over 21 $\mu$m which is relatively long even in Rh-based atoms. This is likely due to the magnetic moments of Rh induced by exchange interaction and the resultant coherent spin wave propagation. These results clearly indicate that B2-ordered FeRh can be used as a material for spin wave transmission bus by integrating with other fascinating magnetic characteristics of FeRh such as electric field induced magnetic phase transition. 


%
%

%

This work was supported in part by JSPS KAKENHI Grant Numbers 26289229, 15H01014, 16K14381, 15H05702, and 15H01998.


\end{document}